\documentclass[aps,floatfix,showpacs,twocolumn,superscriptaddress]{revtex4} 
 
\usepackage{graphicx} 
\usepackage{dcolumn} 
\usepackage{amsmath}

\begin{document} 
 
\preprint{\parbox[t]{45mm}{\small ANL-PHY-10460-TH-2002\\%
                                  MPG-VT-UR 239/02\\%
                                  UNITU-THEP-xx/02}} 
 
\title{
Concerning the quark condensate} 
 
\author{K.\ Langfeld} 
 
\affiliation{Institut f\"ur Theoretische Physik, Universit\"at T\"ubingen, Auf 
der Morgenstelle 14, D-72076 T\"ubingen, Germany} 
 
\author{H.\ Markum} 
 
\affiliation{Atominstitut, Technische Universit\"at Wien, A-1040 
Vienna, Austria} 
 
\author{R.\ Pullirsch} 
 
\affiliation{Institut f\"ur Theoretische Physik, Universit\"at Regensburg, 
D-93040 Regensburg, Germany} 
 
\author{C.D.\ Roberts} 
\affiliation{Physics Division, Bldg 203, Argonne National Laboratory, Argonne 
Illinois 60439-4843} 
\affiliation{Fachbereich Physik, Universit\"at Rostock, D-18051 Rostock, 
Germany} 
 
\author{S.M.\ Schmidt} 
\affiliation{Institut f\"ur Theoretische Physik, Universit\"at T\"ubingen, Auf 
der Morgenstelle 14, D-72076 T\"ubingen, Germany} 
\affiliation{Helmholtz-Gemeinschaft, Ahrstrasse 45, D-53175 Bonn, Germany} 

\begin{abstract} 
\rule{0ex}{3ex} 
A continuum expression for the trace of the massive dressed-quark propagator 
is used to explicate a connection between the infrared limit of the QCD Dirac 
operator's spectrum and the quark condensate appearing in the operator 
product expansion, and the connection is verified via comparison with a 
lattice-QCD simulation.  The pseudoscalar vacuum polarisation provides a good 
approximation to the condensate over a larger range of current-quark masses. 
\end{abstract} 
 
\maketitle 
 
\section{Introduction} 
\label{sec:one} 
Dynamical chiral symmetry breaking (DCSB) is a cornerstone of hadron physics. 
This phenomenon whereby, even in the absence of a current-quark mass, 
self-interactions generate a momentum-dependent running quark mass, $M(p^2)$, 
that is large in the infrared: $M(0) \sim 0.5\,$GeV, but power-law suppressed 
in the ultraviolet \cite{masslessuv}: 
\begin{equation} 
\label{Mchiral} 
M(p^2) \stackrel{{\rm large}-p^2}{=}\, 
\frac{2\pi^2\gamma_m}{3}\,\frac{\left(-\,\langle \bar q q \rangle^0\right)} 
{p^2 \left(\frac{1}{2}\ln\left[\frac{p^2}{\Lambda_{\rm QCD}^2}\right] 
\right)^{1-\gamma_m}}\,, 
\end{equation} 
is impossible in weakly interacting theories.  In Eq.\ (\ref{Mchiral}), 
$\gamma_m=12/(33-2N_f)$ is the mass anomalous dimension, with $N_f$ the 
number of light-quark flavours, and $\langle \bar q q \rangle^0$ is the 
renormalisation-group-invariant vacuum quark condensate \cite{mr97}, to which 
we shall hereafter refer as the OPE condensate.  While Eq.\ (\ref{Mchiral}) 
is expressed in Landau gauge, $\langle \bar q q \rangle^0$ is gauge parameter 
independent.  In the chiral limit the OPE condensate plays a role analogous 
to that played by the renormalisation-group-invariant current-quark mass in 
the massive theory: it sets the scale of the mass function in the 
ultraviolet. 
 
The evolution of the dressed-quark mass-function in Eq.\ (\ref{Mchiral}) to a 
large and finite constituent-quark-like mass in the infrared, $M(0) \sim 
0.5\,$GeV, is a longstanding prediction of Dyson-Schwinger equation (DSE) 
studies \cite{cdragw} that has recently been confirmed in simulations of 
quenched lattice-QCD \cite{latticequark}.  A determination of the OPE 
condensate directly from lattice-QCD simulations must await an accurate 
chiral extrapolation \cite{latticequarkchiral} but DSE models tuned to 
reproduce modern lattice data give \cite{pmqnp2002} $(-\langle \bar q q 
\rangle^0) \sim \Lambda_{\rm QCD}^3$. 
 
Another view of DCSB is obtained by considering the eigenvalues and 
eigenfunctions of the massless Euclidean Dirac operator \cite{fn:Eucl}: 
\begin{equation} 
\gamma\cdot D \, u_n(x) = i \lambda_n \, u_n(x)\,. 
\end{equation} 
The operator is anti-Hermitian and hence the eigenfunctions form a complete 
set, and except for zero modes they occur in pairs: $\{u_n(x), \gamma_5 
u_n(x)\}$, with eigenvalues of opposite sign.  It follows that in an 
external gauge field, $A$, one can write the Green function for a massive 
propagating quark in the form 
\begin{equation} 
S(x,y;A) = \langle q(x) \bar q(y) \rangle_A = \sum_n \frac{u_n(x) \, 
u_n^\dagger(y)}{i \lambda_n + m}\,, 
\end{equation} 
where $m$ is the current-quark mass and, naturally, the eigenvalues depend on 
$A$.  (NB.\ The expectation value denotes a Grassmannian functional integral 
evaluated with a fixed gauge field configuration.)  Assuming, e.g., a lattice 
regularisation, it follows that 
\begin{equation} 
\label{qbqonV} 
\frac{1}{V} \, \int_V d^4x \, \langle \bar q(x) q(x) \rangle_A = - \frac{2 m}{V} 
\, \sum_{\lambda_n>0} \, \frac{1}{\lambda_n^2 + m^2}\,, 
\end{equation} 
where $V$ is the lattice volume \cite{fn:zero}.  One may now define a quark 
condensate as the infinite volume limit of the average in Eq.\ (\ref{qbqonV}) 
over all gauge field configurations: 
\begin{equation} 
\label{BCqbq1} 
\langle 0 | \bar q q | 0 \rangle := \lim_{V\to \infty} \frac{1}{V} \, \int_V 
d^4x \, \left\langle \, \langle \bar q(x) q(x) \rangle_A \,\right\rangle\,. 
\end{equation} 
In the infinite volume limit the operator spectrum becomes dense and Eqs.\ 
(\ref{qbqonV}), (\ref{BCqbq1}) become 
\begin{equation} 
\label{bcone} 
- \langle 0 | \bar q q | 0 \rangle = 2 m \int_0^\infty d\lambda\, 
\frac{\rho(\lambda)}{ \lambda^2 + m^2}\,, 
\end{equation} 
with $\rho(\lambda)$ the spectral density.  This equation expresses an 
assumption that in QCD the full two-point massive-quark Schwinger function, 
when viewed as a function of the current-quark mass, has a spectral 
representation. 
 
It follows formally from Eq.\ (\ref{bcone}) that 
\begin{equation} 
\lim_{m\to 0} \, m \int_0^\infty d\lambda\, 
\frac{\rho(\lambda)}{ \lambda^2 + m^2} = \frac{1}{2} \, \pi \, \rho(0)\,, 
\end{equation} 
and hence one arrives at the chiral limit result 
\begin{equation} 
\label{bcrelation} 
 - \langle 0 | \bar q q | 0 \rangle =  \pi \, \rho(0)\,. 
\end{equation} 
This is the so-called Banks-Casher relation \cite{bankscasher,Marinari}.  It 
has long been advocated as a means by which a quark condensate may be measured 
in lattice-QCD simulations \cite{Marinari} and has been used in analysing 
chiral symmetry restoration at nonzero temperature \cite{Harald} and chemical 
potential \cite{Harald2}, and to explore the connection between magnetic 
monopoles and chiral symmetry breaking in $U(1)$ gauge theory \cite{simon}. 
Much has been learnt \cite{verbaarschot,Jacextras} by exploiting the fact that 
qualitative features of the behaviour of $\rho(\lambda)$ for $\lambda \sim 0$ 
can be understood using chiral random matrix theory; i.e., from considerations 
based solely on QCD's global symmetries. 
 
Our main goal is to explicate a correspondence between the condensate in Eq.\
(\ref{Mchiral}) and that in Eq.\ (\ref{bcrelation}).  In Sec.\ \ref{sec:two}
we discuss the OPE condensate and its connection with QCD's gap equation, and
emphasise that the residue of the lowest-mass pole-contribution to the
flavour-nonsinglet pseudoscalar vacuum polarisation is a direct measure of
the OPE condensate \cite{mrt98}.  A natural ability to express DCSB through
the formation of a nonzero OPE condensate is fundamental to the success of
DSE models of hadron phenomena \cite{reviews}.  In Sec.\ \ref{sec:three} we
carefully define the trace of the massive dressed-quark propagator and use
that to illustrate a connection between $\rho(0)$ and the OPE condensate,
which we verify via comparison with a lattice simulation.  Section
\ref{sec:four} is an epilogue.
 
\section{OPE Condensate} 
\label{sec:two} 
\subsection{Gap and Bethe-Salpeter equations} 
\label{sec:twoA} 
Dynamical chiral symmetry breaking in QCD is readily explored using the DSE 
for the quark self-energy: 
\begin{eqnarray} 
\nonumber 
\lefteqn{S(p)^{-1}  =  Z_2 \,(i\gamma\cdot p + m^{\rm bm})}\\ 
&&  +\, Z_1 \int^\Lambda_q \, g^2 D_{\mu\nu}(p-q) \frac{\lambda^a}{2}\gamma_\mu 
S(q) \Gamma^a_\nu(q,p) \,, \label{gendse} 
\end{eqnarray} 
wherein: $D_{\mu\nu}(k)$ is the renormalised dressed-gluon propagator; 
$\Gamma^a_\nu(q;p)$ is the renormalised dressed-quark-gluon vertex; $m^{\rm 
bm}$ is the $\Lambda$-dependent current-quark bare mass that appears in the 
Lagrangian; and $\int^\Lambda_q := \int^\Lambda d^4 q/(2\pi)^4$ represents a 
translationally-invariant regularisation of the integral, with $\Lambda$ the 
regularisation mass-scale which is removed to infinity as the completion of 
all calculations.  The quark-gluon-vertex and quark wave function 
renormalisation constants, $Z_1(\zeta^2,\Lambda^2)$ and 
$Z_2(\zeta^2,\Lambda^2)$ respectively, depend on the renormalisation point, 
the regularisation mass-scale and the gauge parameter. 
 
If the current-quark mass changes with flavour, then the solution of 
Eq.~(\ref{gendse}) is flavour dependent: 
\begin{eqnarray} 
\nonumber 
 S_f(p)^{-1} & = & i \gamma\cdot p \, A_f(p^2,\zeta^2) + B_f(p^2,\zeta^2) \\ 
& =& \frac{1}{Z_f(p^2,\zeta^2)}\left[ i\gamma\cdot p + M_f(p^2,\zeta^2)\right] 
\,, \label{sinvp} 
\end{eqnarray} 
and is obtained subject to the condition that at some large, spacelike 
$\zeta^2$ 
\begin{equation} 
\label{renormS} \left.S_f(p)^{-1}\right|_{p^2=\zeta^2} = i\gamma\cdot p + 
m_f(\zeta)\,, 
\end{equation} 
where $m_f(\zeta)$ is the renormalised current-quark mass: 
\begin{equation} 
Z_4(\zeta,\Lambda)\,m_f(\zeta) = Z_2(\zeta,\Lambda) \, m_f^{\rm bm}(\Lambda)\,, 
\end{equation} 
with $Z_4$ the renormalisation constant for the scalar part of the quark 
self-energy.  Since QCD is an asymptotically free theory, the chiral limit is 
defined by 
\begin{equation} 
\label{chirallimit} Z_2(\zeta^2,\Lambda^2) \,  m_f^{\rm bm}(\Lambda) \equiv 
0\,, \;\; \Lambda \gg \zeta 
\end{equation} 
and in this case the scalar projection of Eq. (\ref{gendse}) does not exhibit 
an ultraviolet divergence \cite{mrt98,mr97}. 
 
Important in describing chiral symmetry is the axial-vector Ward-Takahashi 
identity: 
\begin{eqnarray} 
\nonumber P_\mu \Gamma_{5\mu}^H(k;P)  & = & {\cal 
S}^{-1}(k_+) \, i\gamma_5\frac{T^H}{2} +  i\gamma_5\frac{T^H}{2} {\cal 
S}^{-1}(k_-) \\ 
& & \nonumber 
- M_{(\zeta)}\,i\Gamma_5^H(k;P) - i\Gamma_5^H(k;P)\,M_{(\zeta)} \,,\\ 
\label{avwti} 
\end{eqnarray} 
where $P$ is the total momentum entering the vertex.  In Eq.\ (\ref{avwti}): 
$M_{(\zeta)}= {\rm diag}[m_u(\zeta),m_d(\zeta),m_s(\zeta)]$, ${\cal S}= {\rm 
diag}[S_u,S_d,S_s]$ and $\{T^H\}$ are flavour matrices, e.g., $T^{\pi^+} = 
\frac{1}{2}\left(\lambda^1 + i \lambda^2\right)$ (we consider SU$_f(3)$ 
because chiral symmetry is unimportant for heavier quarks); 
$\Gamma_{5\mu}^H(k;P)$ is the renormalised axial-vector vertex, which is 
obtained from the inhomogeneous Bethe-Salpeter equation 
\begin{eqnarray} 
\nonumber \lefteqn{\left[\Gamma_{5\mu}^H(k;P)\right]_{tu}  = 
Z_2 \, \left[\gamma_5\gamma_\mu \frac{T^H}{2}\right]_{tu} }\\ 
& + & 
\int^\Lambda_q \, [\chi_{5\mu}^H(q;P)]_{sr} \,K^{rs}_{tu}(q,k;P)\,, 
\label{genave} 
\end{eqnarray} 
where $\chi_{5\mu}^H := {\cal S}(q_+) \Gamma_{5\mu}^H(q;P) {\cal S}(q_-)$, 
$q_\pm = q \pm P/2$, and $K(q,k;P)$ is the fully renormalised quark-antiquark 
scattering kernel; and $\Gamma_{5}^H$ is the pseudoscalar vertex, 
\begin{eqnarray} 
\nonumber \lefteqn{\left[\Gamma_{5}^H(k;P)\right]_{tu}  = 
Z_4\,\left[\gamma_5 \frac{T^H}{2}\right]_{tu} } \\ 
& + & \int^\Lambda_q \, \left[ \chi_5^H(q;P)\right]_{sr} K^{rs}_{tu}(q,k;P)\,, 
\label{genpve} 
\end{eqnarray} 
with $\chi_{5}^H := {\cal S}(q_+) \Gamma_{5}^H(q;P) {\cal S}(q_-)$. 
Multiplicative renormalisability ensures that no new renormalisation 
constants appear in Eqs.~(\ref{genave}) and (\ref{genpve}). 
 
Flavour-octet pseudoscalar bound states appear as coincident pole solutions 
of Eqs.\ (\ref{genave}), (\ref{genpve}), namely, 
\begin{equation} 
\Gamma_{5\mu}^H(k;P) \propto 
\Gamma_{5}^H(k;P) \propto \frac{1}{P^2+m_H^2}\, \Gamma_H(k;P) \,, 
\end{equation} 
where $\Gamma_H$ is the bound state's Bethe-Salpeter amplitude and $m_H$, its 
mass.  (Regular terms are overwhelmed at the pole.)  Consequently, Eq.\ 
(\ref{avwti}) entails \cite{mrt98,mr97} 
\begin{equation} 
\label{gengmor} 
f_H \, m_H^2 = r_H^{(\zeta)}\, {\cal M}_H^{(\zeta)}\,, 
\end{equation} 
where: ${\cal M}_H^{(\zeta)}= {\rm tr}_{\rm flavour}[ 
M_{(\zeta)}\{T^H,(T^H)^{\rm t}]$ is the sum of the constituents' 
current-quark masses (``t'' denotes matrix transpose); and 
\begin{eqnarray} 
f_H P_\mu & = & 
Z_2\int^\Lambda_q\,\frac{1}{2} 
{\rm tr}\left[\left(T^H\right)^{\rm t} \gamma_5 \gamma_\mu 
 \chi_H(q;P)\right]\,, \label{fpidef}\\ 
i r_H^{(\zeta)} & = & Z_4\int^\Lambda_q\,\frac{1}{2} {\rm 
tr}\left[\left(T^H\right)^{\rm t} \gamma_5 \ \chi_H(q;P)\right]\,, 
\label{rHint} 
\end{eqnarray} 
where $\chi_H:= {\cal S}(q_+) \Gamma_H(q;P) {\cal S}(q_-)$ and the 
expressions are evaluated at $P^2+m_H^2=0$. 
 
Equation (\ref{fpidef}) is the pseudovector projection of the meson's 
Bethe-Salpeter wave function evaluated at the origin in configuration space. 
It is the precise expression for the leptonic decay constant.  The 
renormalisation constant, $Z_2(\zeta,\Lambda)$, ensures that the r.h.s.\ is 
independent of: the regularisation scale, $\Lambda$, which may therefore be 
removed to infinity; the renormalisation point; and the gauge parameter. 
Hence it is truly an observable. 
 
Equation (\ref{rHint}) is the pseudoscalar analogue.  Therein the 
renormalisation constant $Z_4(\zeta,\Lambda)$ entails that the r.h.s.\ is 
independent of the regularisation scale, $\Lambda$, and the gauge parameter. It 
also ensures that the $\zeta$-dependence of $r_H^{(\zeta)}$ is precisely that 
required to guarantee the r.h.s.\ of Eq.\ (\ref{gengmor}) is independent of the 
renormalisation point.  (NB.\ $r_H^{(\zeta)}$ is finite, and Eq.\ 
(\ref{gengmor}) valid, for arbitrary values of the current-quark masses 
\cite{heavy,cdrlc01}.) 
 
In the chiral limit the existence of a solution of Eq.\ (\ref{gendse}) with 
$B_0(p^2)\neq 0$; i.e., DCSB, necessarily entails \cite{mrt98} that Eqs.\ 
(\ref{genave}), (\ref{genpve}) exhibit a massless pole solution: the 
Goldstone mode, which is described by 
\begin{eqnarray} 
\label{genpibsa} 
\Gamma_0^g(k;P) & = &  \lambda^g \gamma_5 \bigg[ i E_0(k;P) + 
\gamma\cdot P F_0(k;P) \\ 
\nonumber 
& + &  \gamma\cdot k \,k \cdot P\, G_0(k;P) 
+ \sigma_{\mu\nu}\,k_\mu P_\nu \,H_0(k;P) 
\bigg]\,, 
\end{eqnarray} 
wherein $f_\pi^0 E_0(k;0) = B_0(k^2)$.  (The index ``$0$'' indicates a 
quantity calculated in the chiral limit.)  It follows immediately 
\cite{mrt98} that 
\begin{equation} 
f_\pi^0\, r_0^{(\zeta)} = - \langle \bar q q \rangle_{\zeta}^0= \lim_{\Lambda 
 \to \infty} Z_4(\zeta,\Lambda)\, N_c\, {\rm tr}_D \!  \int^\Lambda_q \! 
 S_{0}(q)\,, 
\label{qbqzeta} 
\end{equation} 
where the trace is only over Dirac indices.  This result and multiplicative 
renormalisability entail 
\begin{equation} 
\frac{\langle \bar q q \rangle_{\zeta}^0}{\langle \bar q q 
\rangle_{\zeta^\prime}^0} = Z_4(\zeta,\zeta^\prime) 
Z_2^{-1}(\zeta,\zeta^\prime) = Z_m(\zeta,\zeta^\prime)\,, 
\end{equation} 
where $Z_m$ is the mass-renormalisation constant.  It is thus apparent that 
the chiral limit behaviour of $r_H^{(\zeta)}$ yields the OPE condensate 
evolved to a renormalisation point $\zeta$. 
 
It is important to recall that the DSEs reproduce every diagram in 
perturbation theory.  Therefore a weak coupling expansion of Eq.\ 
(\ref{gendse}) yields the perturbative series for the dressed-quark 
propagator.  This may be illustrated by the result for the scalar piece of 
the propagator calculated in this way: 
\begin{equation} 
\label{Bf0} 
B_f(p^2) = m_f \left( 1 - \frac{\alpha}{\pi} \ln\left[\frac{p^2}{m_f^2}\right] + 
\ldots \right)\,. 
\end{equation} 
Every term in the series is proportional to the current-quark mass and hence 
a nonzero value of the OPE condensate is impossible in perturbation theory. 
 
\subsection{Pseudoscalar Vacuum Polarisation} 
Consider the colour singlet Schwinger function describing the pseudoscalar 
vacuum polarisation 
\begin{equation} 
\label{Delta5} 
\Delta_5(x) = \langle \bar q(x)\frac{1}{2} \lambda^f\gamma_5 q(x)\, \bar 
q(0)\frac{1}{2} \lambda^g\gamma_5 q(0) \rangle \,, 
\end{equation} 
which can be estimated, e.g., in lattice simulations.  Its renormalised form
can completely be expressed in momentum space using quantities introduced
already:
\begin{equation} 
\label{omega5} 
\omega_5^{fg}(P) = Z_2^2\,{\rm tr}\int^\Lambda_q\!  \frac{1}{2} \lambda^f\gamma_5 
\, S(q_+) \, \Gamma_5^g(q;P) \, S(q_-)\,. 
\end{equation} 
 
Equation (\ref{genpve}) can be rewritten in terms of the fully-amputated 
quark-antiquark scattering amplitude: $M = K + K (SS) K + \ldots$, and in the 
neighbourhood of the lowest mass pole 
\begin{equation} 
\label{Mpole} 
M(q,k;P) = \Gamma_H(q;P) \, \frac{1}{P^2+m_H^2}\, \bar\Gamma(k;-P) + 
R(q,k;P)\,, 
\end{equation} 
where $R$ is regular in this neighbourhood. 
 
Assuming SU$_f(3)$ flavour symmetry, substituting Eq.\ (\ref{Mpole}) into 
Eq.\ (\ref{omega5}) gives 
\begin{equation} 
\omega_5^{fg}(P) = \delta^{fg}\, \frac{1}{P^2+m_H^2}\, Z_m^{-2} \, r_H^2 + 
\ldots 
\end{equation} 
(the ellipsis denotes terms regular in the pole's neighbourhood).  It follows 
that the large-$x^2$ behaviour of 
\begin{equation} 
m^2_{\rm bm}\, \Delta_5(x) 
\end{equation} 
is a measure of the renormalisation-group-invariant 
\begin{equation} 
\left[ m(\zeta) \, r_H^{(\zeta)} \right]^2 
\end{equation} 
that appears in Eq.\ (\ref{gengmor}).  Hence the correlator in Eq.\ 
(\ref{Delta5}) provides a direct means of estimating the OPE condensate in 
lattice simulations \cite{latticeOPE}, one whose ultraviolet behaviour ensures 
a well-defined and calculable evolution under the renormalisation group for any 
value of the current-quark mass.  (NB.\ $f_\pi$ can similarly be extracted from 
the axial-vector correlator analogous to Eq.\ (\ref{Delta5}).)  The model of 
Ref.\ \cite{pmspectra2} yields a meson mass trajectory via Eq.\ (\ref{gengmor}) 
that provides a qualitative and quantitative understanding of recent quenched 
lattice simulations \cite{cdrlc01}. 
 
\section{Banks-Casher Relation} 
\label{sec:three} 
\subsection{Continuum analysis} 
It is readily apparent that Eq.\ (\ref{bcone}) is meaningless as written: 
dimensional counting reveals the r.h.s.\ has mass-dimension three and since 
$\lambda$ will at some point be greater than any relevant internal scale, the 
integral must diverge as $\Lambda^2$, where $\Lambda$ is the regularising 
mass-scale. 
 
To learn more, consider the trace of the unrenormalised massive dressed-quark 
propagator: 
\begin{equation} 
\label{defsigmam} 
\tilde \sigma(m) := N_c\,{\rm tr}_D\int_p^\Lambda\! \tilde S_m(p) \,, 
\end{equation} 
evaluated at a fixed value of the regularisation scale, $\Lambda$.  This
Schwinger function can be identified with the l.h.s.\ of Eq.\ (\ref{bcone}).
Furthermore, assume that $\tilde\sigma(m)$ has a spectral representation,
since this is the essence of the Banks-Casher relation:
\begin{equation} 
\label{defsigmamspectral} 
\tilde \sigma(m) := 2 m \, \int_0^\Lambda \! d \lambda \frac{\tilde 
\rho(\lambda)}{\lambda^2 + m^2}\,, 
\end{equation} 
where $m = m^{\rm bm}(\Lambda)$.  Equation (\ref{defsigmamspectral}) entails 
\begin{equation} 
\label{discty} 
\tilde \rho(\lambda) = \frac{1}{2 \pi} \lim_{\eta\to 0^+} \left[ \, \tilde 
\sigma(i \lambda + \eta) - \tilde \sigma(i \lambda - \eta)\, \right]\,. 
\end{equation} 
 
The content and meaning of this sequence of equations is well illustrated by 
inserting the free quark propagator in Eq.\ (\ref{defsigmam}).  The integral 
thus obtained is readily evaluated using dimensional regularisation: 
\begin{equation} 
\tilde \sigma_{\rm free}(m) = \frac{N_c}{4\pi^2} \, m^3 \left[ 
\ln\frac{m^2}{\zeta^2}+ \frac{1}{\varepsilon} + \gamma - \ln 4\pi \right]. 
\end{equation} 
With Eq.\ (\ref{discty}) the regularisation dependent terms cancel and one 
obtains 
\begin{equation} 
\label{rhopert} 
\tilde \rho(\lambda) = \frac{N_c}{4\pi^2} \, \lambda^3\,. 
\end{equation} 
 
The one-loop contribution to $\tilde \rho(\lambda)$ has been evaluated using
the same procedure \cite{rholambda1}.  It is also proportional to $\lambda^3$
and arises from the $m^3 \ln m^2/\zeta^2$ terms in $\tilde\sigma(m)$.  In
fact, every term obtainable in perturbation theory is proportional to
$\lambda^3$, for precisely the same reason that each term in the perturbative
expression for the scalar part of the quark propagator is proportional to
$m$, see Eq.\ (\ref{Bf0}).  Hence, at every order in perturbation theory,
\begin{equation} 
\tilde \rho(\lambda=0) = 0 
\end{equation} 
and $\langle 0 | \bar q q | 0 \rangle =0$.  A nonzero value of $\rho(0)$ is 
plainly an essentially nonperturbative effect. 
 
A precise analysis requires that attention be paid to renormalisation.
Consider then the gauge-parameter-independent trace of the renormalised quark
propagator evaluated at a fixed value of the regularisation scale:
\begin{eqnarray} 
\nonumber 
\sigma(m;\zeta) & := & -\lim_{x\to 0} \langle \bar q(x) q(0) \rangle \\ 
& = & Z_4(\zeta,\Lambda)\, N_c\,{\rm tr}_D\int_p^\Lambda\! S_m(p;\zeta) \,, 
\label{sigmam} 
\end{eqnarray} 
where the argument remains $m=m^{\rm bm}(\Lambda)$, which is permitted 
because $m^{\rm bm}(\Lambda)$ is proportional to the 
renormalisation-point-independent current-quark mass.  The renormalisation 
constant $Z_4$ vanishes logarithmically with increasing $\Lambda$ and hence 
one still has $\sigma(m) \sim \Lambda^2\, m^{\rm bm}(\Lambda)$.  However, 
using Eq.\ (\ref{qbqzeta}) it is clear that for any finite but large value of 
$\Lambda$ and tolerance $\delta$, it is always possible to find 
$m_\delta^{\rm bm}(\Lambda)$ such that 
\begin{equation} 
\label{sigmabound} 
\sigma(m;\zeta) + \langle \bar q q \rangle^0_\zeta < \delta\,,\; \forall
m^{\rm bm}< m_\delta^{\rm bm}.
\end{equation} 
 
\begin{figure}[t] 
\centerline{\resizebox{0.45\textwidth}{!}{\includegraphics{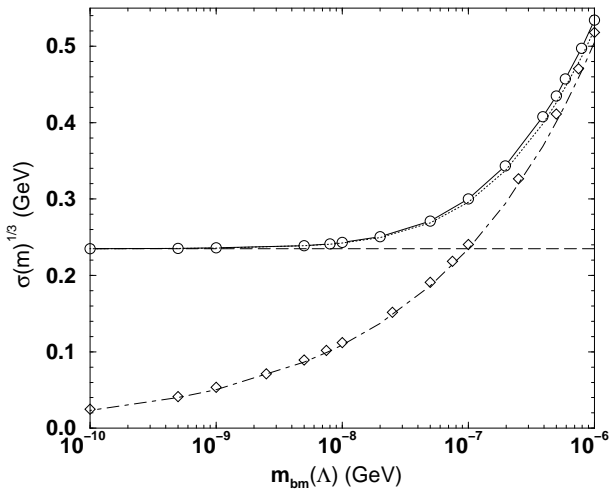}}} 
 
\caption{\label{fig1} Circles/solid-line: $\sigma(m)^{1/3}$ in Eq.\ 
(\protect\ref{sigmam}) as a function of the current-quark bare-mass, evaluated 
using the dressed-quark propagator obtained in the model of Ref.\ 
\protect\cite{mr97}; dashed line: the model's value of $(-\langle \bar q q 
\rangle^0_{\zeta=1\,{\rm GeV}}) = (0.24\,{\rm GeV})^3$; and dotted line: Eq.\ 
(\protect\ref{sigmamexpected}).  Diamonds: $\sigma(m)^{1/3}$ evaluated in a 
non-confining version of the model; dot-dashed line: Eq.\ 
(\protect\ref{sigmamexpected}) with $\langle \bar q q \rangle^0 \equiv 0$.} 
\end{figure}

This is true in QCD.  It can be illustrated using the DSE model of Ref.\ 
\cite{mr97}, which preserves the one-loop renormalisation group properties of 
QCD.  In Fig.\ \ref{fig1} we plot $\sigma(m,\zeta)$, evaluated using a hard 
cutoff, $\Lambda$, on the integral in Eq.\ (\ref{sigmam}), calculated with the 
massive dressed-quark propagators obtained by solving the gap equation as 
described in the appendix.  Since Eq.\ (\ref{sigmabound}) specifies the domain 
on which the value of $\sigma(m;\zeta)$ is determined by nonperturbative 
effects, one anticipates 
\begin{equation} 
\label{epsilondomain} 
m_\delta^{\rm bm}(\Lambda) \approx -\langle \bar q q \rangle^0 / 
\Lambda^2 \sim 10^{-9} 
\end{equation} 
for $\Lambda = 2.0\,$TeV in QCD where $|\langle \bar q q \rangle^0| \sim 
\Lambda_{\rm QCD}^3$, an estimate confirmed in Fig.\ \ref{fig1}. 
 
The dotted line in Fig.\ \ref{fig1} is 
\begin{eqnarray} 
\nonumber \lefteqn{\sigma(m,\zeta) = -\langle \bar q q \rangle^0_\zeta \, 
\frac{2}{\pi} \arctan \frac{\Lambda}{m} 
}\\ 
&+ &  Z_4(\zeta,\Lambda)\, \frac{N_c}{4\pi^2} \, m\left[ \Lambda^2 - m^2\ln 
\left(1+\Lambda^2 / m^2\right)\right] \,. 
\label{sigmamexpected} 
\end{eqnarray} 
(NB.\ We used the one-loop formula: $Z_4(\zeta,\Lambda)=[
\alpha(\Lambda)/\alpha(\zeta)]^{\gamma_m}$, for the numerical comparison.)
The difference between Eq.\ (\ref{sigmamexpected}) and the curve is of O$(
\alpha(\Lambda) \, m(\Lambda)\, \Lambda^2) $ because the DSE model
incorporates QCD's one-loop behaviour.  In Fig.\ \ref{fig1} we also plot
$\sigma(m,\zeta)$ obtained in the absence of confinement, in which case
\cite{hawes} $\langle \bar q q \rangle^0\equiv 0$, as is apparent.
 
The discussion establishes that $\sigma(m,\zeta)$ has a regular chiral limit in 
QCD and is a monotonically increasing convex-up function.  It follows that 
$\sigma(m,\zeta)$ has a spectral representation: 
\begin{equation} 
\sigma(m,\zeta) = 2 m \int_0^\Lambda \frac{\rho(\lambda)}{\lambda^2 + m^2}\,. 
\end{equation} 
This lays the vital plank in a veracious connection between the condensates in 
Eqs.\ (\ref{Mchiral}) and (\ref{bcrelation}).  On the domain specified by Eq.\ 
(\ref{epsilondomain}), the behaviour of $\sigma(m,\zeta)$ in Eq.\ 
(\ref{sigmam}) is given by Eq.\ (\ref{sigmamexpected}), which yields, via Eq.\ 
(\ref{discty}), 
\begin{equation} 
\label{finalrho} 
\pi\, \rho(\lambda) = -\langle \bar q q \rangle^0_\zeta + Z_4(\zeta,\Lambda)\, 
\frac{N_c}{4\pi}  \, \lambda^3 + \ldots 
\end{equation} 
where the ellipsis denotes contributions from the higher-order terms implicit 
in Eq.\ (\ref{sigmamexpected}). 
 
\subsection{Comparison with a lattice-QCD simulation} 
\begin{figure}[t] 
\centerline{\resizebox{0.45\textwidth}{!}{\includegraphics{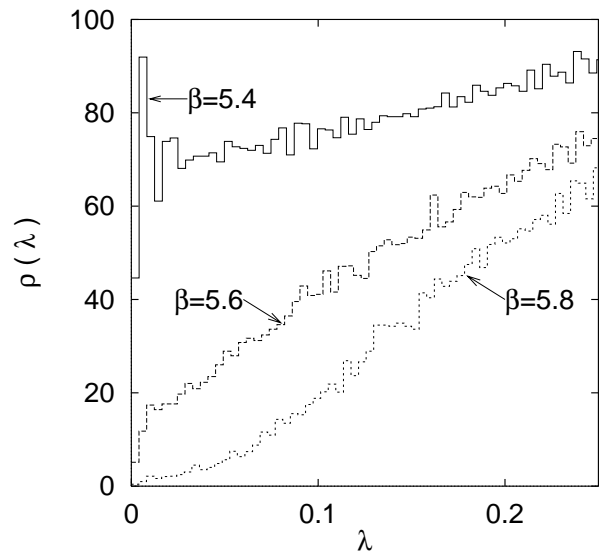}}} 
\caption{\label{latticerho} Spectral density of the staggered Dirac operator in 
quenched $SU(3)$ gauge theory calculated on a $4^4$-lattice.  (Measured in 
units of the lattice spacing.)  The deconfinement transition takes place at 
$\beta \gtrsim 5.6$. } 
\end{figure}

In Fig.\ \ref{latticerho} we plot the spectral density of the staggered Dirac 
operator in quenched $SU(3)$ gauge theory calculated with $3000$ configurations 
obtained on a $V = 4^4$-lattice, in the vicinity of the deconfining phase 
transition at $\beta \gtrsim 5.6$.  Details of the simulation are given in 
Ref.\ \cite{Harald}.  Dimensioned quantities are measured in units of $1/a$, 
where $a$ is the lattice spacing, and it is $\rho(\lambda)/V$ that should be 
compared with the continuum spectral density. 
 
While the effect of finite lattice volume is apparent in Fig.\ \ref{latticerho} 
for $\lambda a \gtrsim 0.1$, the behaviour at small $\lambda a$ is 
qualitatively in agreement with Eq.\ (\ref{finalrho}) and Fig.\ \ref{fig1}: a 
nonzero OPE condensate dominates the Dirac spectrum in the confined domain; and 
it vanishes in the deconfined domain whereupon $\rho(0)=0$ and the perturbative 
evolution, Eq.\ (\ref{rhopert}), is manifest. 
 
To be more quantitative, we note that at $\beta = 5.4$, $\rho(0)a^3 \approx 
70$, so that 
\begin{equation} 
\label{pirho0lattice} \pi \rho(0) a^3/V \approx (0.95)^3. 
\end{equation} 
The value of the lattice spacing was not measured in Ref.\ \cite{Harald} but 
one can nevertheless assess the scale of Eq.\ (\ref{pirho0lattice}) by 
supposing $a \sim 0.3\, {\rm fm} \sim 0.3/ \Lambda_{\rm QCD}$, a value typical 
of small couplings, $\beta$, wherewith the r.h.s.\ is $\sim (3 \Lambda_{\rm 
QCD})^3\!$. This is too large but not unreasonable given the parameters of the 
simulation, its errors and the systematic uncertainties in our estimate. One 
can also fit the lattice data at $\beta = 5.8$, whereby one finds 
$\rho(\lambda) \propto \lambda^3 $ on $\lambda< 0.1$ but with a proportionality 
constant larger than that anticipated from perturbation theory; viz.\ Eq.\ 
(\ref{finalrho}). Some mismatch is to be expected because at $\beta=5.8$ one 
has only just entered the deconfined domain and close to the transition 
boundary nonperturbative effects are still material, as seen, e.g., in the 
heavy-quark potential and equation of state \cite{edwinechaya}. It is a modern 
challenge to determine those gauge couplings and lattice parameters for which 
the data are quantitatively consistent with Eq.\ (\ref{finalrho}). 
 
\section{Epilogue} 
\label{sec:four} 
We verified that the gauge-invariant trace of the massive dressed-quark 
propagator possesses a spectral representation when considered as a function of 
the current-quark mass.  This is key to establishing that the OPE condensate, 
which sets the ultraviolet scale for the momentum-dependence of the trace of 
the dressed-quark propagator, does indeed measure the density of far-infrared 
eigenvalues of the gauge-averaged massless Dirac operator, \textit{\`{a} la} 
the Banks-Casher relation.  This relation is intuitively appealing because a 
measurable accumulation of eigenvalues of the massless Dirac operator at 
zero-virtuality expresses a mass gap in its spectrum. 
 
In practice, there are three main parameters in a simulation of lattice-QCD: 
the lattice volume, characterised by a length $L$; the lattice spacing, $a$; 
and the current-quark mass, $m$.  So long as the lattice size is large compared 
with the current-quark's Compton wavelength; viz., $L\gg 1/m$, then dynamical 
chiral symmetry breaking can be expressed in the simulation. Supposing that to 
be the case then, as we have explicated, so long as the lattice spacing is 
small compared with the current-quark's Compton wavelength; i.e., $a\ll 1/m\ll 
L $, 
\begin{equation} 
\pi \, \rho(0)  \approx-\langle \bar q q \rangle^0_{1/a} \,, 
\end{equation} 
where the r.h.s.\ is the scale-dependent OPE condensate [$\zeta = \Lambda = 
1/a$ in Eq.\ (\ref{finalrho})]. 
 
In our continuum analysis we found that one requires $a m \lesssim (a 
\Lambda_{\rm QCD})^3$ if $\rho(\lambda= 0)$ is to provide a veracious 
estimate of the OPE condensate.  The residue at the lowest-mass pole in the 
flavour-nonsinglet pseudoscalar vacuum polarisation provides a measure of the 
OPE condensate that is accurate for larger current-quark masses. 
 
\begin{acknowledgments} 
We benefited from interactions with M.A.\ Pichowsky and P.C.\ Tandy.  This 
work was supported by: Deutsche For\-schungs\-ge\-mein\-schaft, under 
contract no.\ Ro~1146/3-1; the Department of Energy, Nuclear Physics 
Division, under contract no.~\mbox{W-31-109-ENG-38}; the National Science 
Foundation under grant no.\ INT-0129236; and benefited from the resources of 
the National Energy Research Scientific Computing Center. 
\end{acknowledgments} 
 
\appendix 
\section{Model Gap Equation} 
The gap equation's kernel is built from a product of the dressed-gluon 
propagator and dres\-sed-quark-gluon vertex.  It can be calculated in 
perturbation theory but that is inadequate for the study of intrinsically 
nonperturbative phenomena.  To make model-independent statements about DCSB 
one must employ an alternative systematic and chiral symmetry preserving 
truncation scheme. 
 
The leading order term in one such scheme \cite{truncscheme} is the 
renormalisation-group-improved rainbow truncation of the gap equation 
$(Q=p-q)$: 
\begin{eqnarray} 
\nonumber \lefteqn{S(p)^{-1} = Z_2 \,(i\gamma\cdot p + m^{\rm bm})}\\ 
& +&  \int^\Lambda_q \, 
{\cal G}(Q^2) D_{\mu\nu}^{\rm free}(Q) \frac{\lambda^a}{2}\gamma_\mu S(q) 
\frac{\lambda^a}{2}\gamma_\nu \,. \label{modeldse} 
\end{eqnarray} 
The ultraviolet ($Q^2 \gtrsim 1\,$GeV$^2)$ behaviour of ${\cal G}(Q^2)$ in 
Eq.\ (\ref{modeldse}) is fixed by the known behaviour of the quark-antiquark 
scattering kernel \cite{mr97}.  The form of that kernel on the infrared 
domain is currently unknown and a model is employed to complete the 
specification of the kernel.  An efficacious form is \cite{mr97} 
\begin{eqnarray} 
\nonumber \lefteqn{ \frac{{\cal G}(Q^2)}{Q^2}  =  8\pi^4 D \delta^4(k) + 
\frac{4\pi^2}{\omega^6} D Q^2 {\rm e}^{-Q^2/\omega^2} } \\ 
&& + 4\pi\,\frac{ \gamma_m 
\pi} {\frac{1}{2} \ln\left[\tau + \left(1 + Q^2/\Lambda_{\rm 
QCD}^2\right)^2\right]} {\cal F}(Q^2) \,, 
\label{gk2} 
\end{eqnarray} 
where: ${\cal F}(Q^2)= [1 - \exp(-Q^2/[4 m_t^2])]/Q^2$, $m_t$ $=$ $0.5\,$GeV; 
$\tau={\rm e}^2-1$; $\gamma_m = 12/(33-2 N_f)$, $N_f=4$; and $\Lambda_{\rm 
QCD} = \Lambda^{(4)}_{\overline{\rm MS}}=0.234\,$GeV.  The true parameters in 
Eq.\ (\ref{gk2}) are $D$ and $\omega$, however, they are not independent: in 
fitting, a change in one is compensated by altering the other, with fitted 
observables changing little along a trajectory $\omega \,D = (0.6\,{\rm 
GeV})^3$.  Herein we used 
\begin{equation} 
D= (0.884\,{\rm GeV})^2\,, \; \omega=0.3\,{\rm GeV}. 
\end{equation} 
A non-confining model is obtained with $D=0$. 
 
Equation (\ref{modeldse}), is readily solved for the dressed-quark 
propagator, with the renormalisation constants fixed via Eq.\ 
(\ref{renormS}).  That solution provides the elements used in the 
illustration of Sec.\ \ref{sec:three}. 
 
\vspace*{\fill} 
 


\begin{thebibliography}{00}
\bibitem{fn:zero} In deriving Eq.\ (\protect\ref{qbqonV}), zero modes have 
been neglected, which is justified under broad conditions 
\protect\cite{leutwyler}.
\end{thebibliography}
\end{document}